\documentclass[reprint,amsmath,amssymb,aps,pra,nofootinbib]{revtex4-2}

\usepackage[T2A]{fontenc}
\usepackage[utf8]{inputenc}
\usepackage[russian,english]{babel}
\usepackage[margin=1.8cm]{geometry}
\usepackage[unicode, pdftex]{hyperref}
\usepackage{hypcap}
\usepackage{fancyhdr}
\usepackage{ragged2e} 
\usepackage{amsmath}
\usepackage{amssymb}
\usepackage{amsfonts}
\usepackage{graphicx}
\usepackage{xcolor}
\usepackage{booktabs}
\usepackage{indentfirst}
\usepackage{array}
\usepackage{orcidlink}
\usepackage{marvosym}
\usepackage{bbm}

\definecolor{ugrey}{HTML}{666666}

\renewcommand{\leq}{\leqslant}
\renewcommand{\geq}{\geqslant}

\newcommand{\T}{^{\textnormal{T}}}
\newcommand{\sub}[2]{#1_{\textnormal{#2}}}
\newcommand{\grey}[1]{\textcolor{ugrey}{#1}}

\newcommand{\ket}[1]{\left| #1 \right\rangle}
\DeclareDocumentCommand{\bk}{m o m}{
    \IfNoValueTF{#2}{\langle #1 | #3 \rangle}{\langle #1 | #2 | #3 \rangle}
}

\newcommand{\F}{\mathbb{F}}
\newcommand{\Z}{\mathbb{Z}}

\newcommand{\tca}{\grey{\textsuperscript{[\citenum{ruiz_2025},\citenum{chen_2025},\citenum{cenk_2010}]\phantom{,\citenum{vandaele_2025a}}}}}
\newcommand{\tcc}{\grey{\textsuperscript{[\citenum{ruiz_2025},\citenum{vandaele_2025a},\citenum{chen_2025},\citenum{cenk_2010}]}}}
\newcommand{\tce}{\grey{\textsuperscript{[\citenum{cenk_2010}]\phantom{,\citenum{ruiz_2025},\citenum{vandaele_2025a},\citenum{chen_2025}}}}}
\newcommand{\tcf}{\phantom{\textsuperscript{[\citenum{cenk_2010}]}}}

\newcommand{\ttca}{\grey{\textsuperscript{[\citenum{ruiz_2025}]\phantom{,\citenum{vandaele_2025a}}}}}
\newcommand{\ttcb}{\grey{\textsuperscript{[\citenum{vandaele_2025a}]\phantom{,\citenum{ruiz_2025}}}}}
\newcommand{\ttcc}{\grey{\textsuperscript{[\citenum{ruiz_2025},\citenum{vandaele_2025a}]}}}

\newcommand{\tdagl}{$^{\dag}$}
\newcommand{\tdago}{\phantom{$^{\dag}$}}
\newcommand{\tddgl}{$^{\ddag}$}
\newcommand{\tddgo}{\phantom{$^{\ddag}$}}
\newcommand{\tstrl}{$^*$}
\newcommand{\tstro}{\phantom{$^*$}}
\newcommand{\tatqo}{\phantom{\textsuperscript{\cite{ruiz_2025}}}}
\newcommand{\tvvvo}{\phantom{\textsuperscript{\cite{vandaele_2025a}}}}


\makeatletter
\renewcommand\frontmatter@abstractheading{%
  \begingroup\centering\normalfont\normalsize\bfseries
  \abstractname\par\endgroup\vspace{0.5em}}
\makeatother

\usepackage{titlesec}

\titleformat{\section}[block]{\normalfont\normalsize\scshape\centering}{}{0em}{}
\titleformat{\subsection}[block]{\normalfont\small\bfseries\centering}{}{0em}{}
\titlespacing*{\section}{0pt}{\baselineskip}{0.5\baselineskip}
\titlespacing*{\subsection}{0pt}{\baselineskip}{0.25\baselineskip}

\fancypagestyle{plain}{%
  \fancyhf{}
  \fancyfoot[C]{\thepage}

}

\begin{document}

\setlength{\abovedisplayskip}{3pt}
\setlength{\abovedisplayshortskip}{3pt}
\setlength{\belowdisplayskip}{3pt}
\setlength{\belowdisplayshortskip}{3pt}
\setlength{\headheight}{13pt}
\setlength{\parskip}{0pt}

\title{Tensor Decomposition for Non-Clifford Gate Minimization}

\author{
	Kirill Khoruzhii\orcidlink{0000-0003-4689-3812}$^{1,\text{\Letter}}$,
	Patrick Gelß\orcidlink{0000-0002-3645-9513}$^{1}$,
    Sebastian Pokutta\orcidlink{0000-0001-7365-3000}$^{1,2}$
}
\affiliation{
	$^1$Zuse Institute Berlin, Berlin, Germany\\
    $^2$Technische Universität Berlin, Germany
}

\begin{abstract}
Fault-tolerant quantum computation requires substantial resources for non-Clifford gates. Their count also enters exponentially into the runtime of several classical simulation methods. We present algebraic methods for minimizing the non-Clifford $CCZ$ (Toffoli) count in $\{\mathrm{CNOT},CCZ\}$ phase circuits using tensor decomposition over $\F_2$. On standard cubic benchmarks, these methods match or improve prior automated Toffoli counts, with most instances optimized in under a minute on a single CPU. With additional computation, the decompositions provide effective initializations for existing $T$-count methods, improving $T$ counts on several circuits. We also present algebraic methods for CNOT-count optimization at fixed Toffoli count. This gives competitive ancilla-free finite-field multiplication circuits and reduces the CNOT count of XOR AES S-box circuits.
\end{abstract}

\maketitle
\thispagestyle{fancy}

\begingroup
\renewcommand\thefootnote{}
\footnotetext{\textsuperscript{\Letter}\,Corresponding author: khoruzhii@zib.de}
\endgroup

\begin{figure*}
    \centering
    \includegraphics{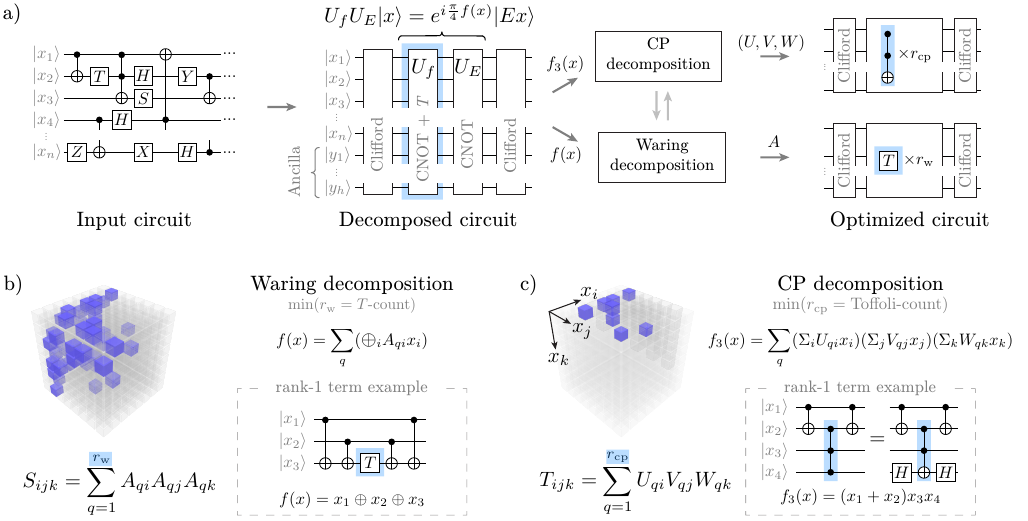}
    \caption{\justifying
    \textbf{Non-Clifford gate minimization as tensor decomposition.}
    a) Overview. Up to Clifford gates, any Clifford+$T$ circuit can be reduced to a diagonal phase operator $U_f$; $U_E$ denotes the accompanying linear map with matrix $E$. Reducing non-Clifford cost is then recast as a tensor-decomposition problem: a Waring decomposition can be applied to optimize the $T$ count of the full phase polynomial $f(x)$, while a canonical polyadic (CP) decomposition targets its cubic part $f_3(x)$ to reduce the number of $CCZ$/Toffoli gates. A CP decomposition can also initialize a subsequent Waring decomposition.
    b) Waring decomposition. The symmetric order-3 signature tensor $S$ is written as a sum of $r_w$ rank-1 terms with factor matrix $A$, giving a realization with $r_w$ $T$ gates. The inset shows one term: compute a parity, apply $T$, and uncompute.
    c) CP decomposition. The cubic tensor $T$ is expressed as a sum of $r_{\mathrm{cp}}$ rank-1 outer products with factor matrices $U,V,W$, corresponding to $r_{\mathrm{cp}}$ $CCZ$/Toffoli gates acting on parities. Two Hadamard ($H$) gates interconvert $CCZ$ and Toffoli, as shown in the inset.
    Colored cubes mark nonzero tensor entries; blue shading highlights non-Clifford gates and their counts.
    }
    \label{fig:workflow}
\end{figure*}

\section{Introduction} \label{sec:introduction}

Fault-tolerant quantum computation relies on error correction schemes that impose different costs on different gate types. Non-Clifford gates such as Toffoli typically have a higher implementation cost than Clifford gates, often requiring magic state distillation~\cite{heyfron_2018,gidney_2019,ruiz_2025}. The relative costs depend on the fault-tolerant implementation, and advances in magic-state preparation can reduce the non-Clifford overhead~\cite{gidney_2024,chen_2026}. Even when CNOT gates are cheaper individually, their total contribution to the circuit cost can be comparable to that of Toffoli gates~\cite{yosri_2026}. We prioritize non-Clifford gate minimization and then optimize the CNOT-count at fixed non-Clifford count, treating single-qubit Clifford gates as free. Non-Clifford gate minimization also benefits classical simulation: known methods achieve runtime polynomial in the number of qubits and Clifford gates but exponential in the non-Clifford gate count~\cite{aaronson_2004,bravyi_2019}.

Phase-polynomial methods connect non-Clifford gate minimization
to tensor decomposition. In Clifford+$T$ circuits, minimizing
the number of non-Clifford $T$ gates is related to decoding
Reed--Muller codes and to symmetric (Waring) decomposition
of order-3 tensors~\cite{amy_2019}. The TODD algorithm~\cite{heyfron_2018} and its FastTODD variant~\cite{vandaele_2025a} operate on CNOT+$T$ subcircuits isolated via Hadamard gadgetization, seeking lower-rank decompositions of the resulting signature tensor.

Clifford+Toffoli provides an alternative gate set for real quantum computation, including reversible arithmetic~\cite{shi_2003}. Since Toffoli and $CCZ$ differ only by Hadamard conjugation, we refer to $CCZ$-count minimization as Toffoli-count minimization. With dedicated $CCZ$ magic state factories~\cite{gidney_2019}, direct minimization of $CCZ$ gates becomes a natural optimization target. AlphaTensor-Quantum~\cite{ruiz_2025} applies deep reinforcement learning to tensor decomposition, rewarding decompositions with Toffoli gadgets. However, its reported training used thousands of TPU actors and took days, limiting accessibility. An independent reusability study~\cite{zen_2026} further highlighted these limitations.

We focus on $n$-qubit diagonal phase operators whose non-Clifford part is realized by $CCZ$ gates, with CNOTs computing the required parities. Bilinear circuits also enter this setting through Hadamard conjugation on the accumulator. Here the non-Clifford content is captured by a cubic polynomial over $\F_2$, and the lower-degree phase is Clifford.

In this work, we develop algebraic methods for Toffoli-count minimization with modest computational requirements; most circuits complete within a minute on a single CPU. Our approach builds on two classical problems in algebra: algebraic thickness, well-studied in cryptographic contexts~\cite{carlet_2002}, and canonical polyadic (CP) decomposition of order-3 tensors, central to fast matrix multiplication~\cite{fawzi_2022,kauers_2023}. When CNOT and $CCZ$ gates are separated into layers, Toffoli-count minimization reduces to algebraic thickness, with a search over $O(2^{n^2})$ basis changes. Allowing CNOTs and $CCZ$ gates to interleave leads to the more general CP formulation and expands the search space to $O(2^{n^3})$. We use basis change optimization (BCO) for the layered case. To find decompositions below algebraic thickness, we develop symplectic Gaussian elimination (SGE) and flip graph search (FGS)~\cite{kauers_2023,moosbauer_2025,chen_2025,khoruzhii_2025}. BCO and SGE are greedy methods with low computational cost; FGS performs random walks on the space of decompositions and is substantially more expensive. For most benchmark circuits, BCO+SGE alone suffices to match or improve prior automated results.

A tensor decomposition fixes the Toffoli count but leaves freedom in the CNOT realization. Cubic-phase routing (CPR) jointly routes the three parities required by each $CCZ$ gate. Quadratic-phase routing (QPR) uses the same route and one-qubit gate insertions to implement the quadratic phase with fewer additional two-qubit gates. We count each $CZ$ gate as one CNOT, since they differ only by Hadamard conjugation.

For bilinear circuits such as $\mathrm{GF}(2^p)$ multiplication, the phase polynomial admits a structured factorization that reduces the search space. Nevertheless, finite field multiplication remains challenging for automated methods: our bilinear search matches or improves prior automated Toffoli counts~\cite{ruiz_2025} but remains above the bounds of known constructions~\cite{cenk_2010} for some field sizes. Applying CPR to decompositions from these constructions provides ancilla-free multiplication circuits with competitive Toffoli and CNOT counts. We also use CPR and QPR to optimize arithmetic blocks in existing XOR AES S-box constructions~\cite{li_2023,chen_2025b}, obtaining S-box circuits with reduced CNOT counts.

Finally, the CP decompositions generated by FGS provide effective initializations for $T$-count optimization with TODD~\cite{heyfron_2018,vandaele_2025a}. Expanding their $CCZ$ terms into $T$ gates gives diverse starting points leading to improved $T$ counts on several benchmark circuits.

We evaluate our methods on planted benchmarks with known structure and on standard circuit families from prior work~\cite{heyfron_2018,ruiz_2025,vandaele_2025a}.

\begin{figure*}
    \centering
    \includegraphics{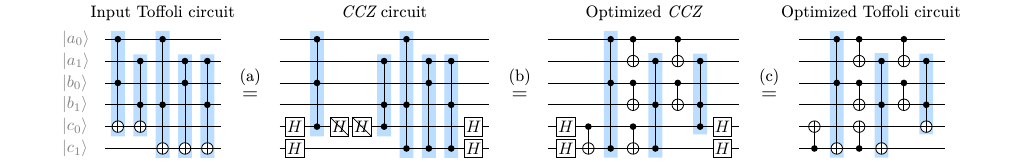}
\caption{\justifying
\textbf{Bilinear circuit optimization.}
The example shows multiplication over the four-element field $\mathrm{GF}(2^2)$, with input registers $a,b$ and accumulator $c$.
(a) Hadamard ($H$) conjugation on $c$ converts Toffoli gates to $CCZ$ gates; adjacent $H$ gates cancel.
(b) A low-rank canonical polyadic (CP) decomposition of the bilinear tensor reduces the $CCZ$ count from 5 to 3.
(c) Inverse Hadamard conjugation restores Toffoli form.
Blue shading marks Toffoli and $CCZ$ gates.
}
    \label{fig:gf-mult}
\end{figure*}

\section{Results} \label{sec:results}

\subsection{Tensor decomposition} \label{sec:results:a}

Any Clifford+$T$ circuit can be reduced to a form in which all non-Clifford operations are collected into a single diagonal phase operator $U_f$, sandwiched between Clifford layers (Fig.~\ref{fig:workflow}a). The standard technique, Hadamard gadgetization~\cite{heyfron_2018}, replaces each internal Hadamard with a gadget consuming one ancilla qubit, expanding an $n$-qubit circuit to $n+h$ qubits where $h$ is the number of internal Hadamards.

The phase operator acts as $U_f \ket{x} = e^{i\frac{\pi}{4} f(x)} \ket{x}$, where $f\colon \F_2^n \to \Z_8$ is a polynomial of degree at most~3:
\begin{equation}
    f = f_1 + 2\, f_2 + 4\, f_3 + \varphi.
    \label{eq:weighted-poly}
\end{equation}
Here $f_1$, $f_2$, $f_3$ are homogeneous polynomials of degree 1, 2, and~3. Their monomials correspond to powers of $T$, $CS$, and $CCZ$ gates, respectively~\cite{heyfron_2018}. Since two identical gates of each type compose to a Clifford gate
\begin{equation}
    T^2 = S,\ CS^2 = CZ,\ CCZ^2 = I,
\end{equation}
we take $f_1,f_2,f_3$ over $\F_2$ and collect the remaining diagonal Clifford phase in~$\varphi$.

The total non-Clifford cost is
\begin{equation}
    c_1 |f_1| + c_2 |f_2| + c_3 |f_3|,
    \label{eq:cost}
\end{equation}
where $|f|$ denotes the number of monomials and the weights $(c_1, c_2, c_3)$ depend on the target cost model: $(1, 3, 7)$ for unitary $T$-gate decomposition, or $(1, 2, 2)$ when dedicated $CCZ$ factories are considered~\cite{ruiz_2025,gidney_2019}.

For circuits with $f_1 = 0$ and $f_2 = 0$, which cover nearly all standard benchmarks~\cite{ruiz_2025,vandaele_2025a,heyfron_2018}, the phase operator takes the form $U_f \ket{x} = e^{i\frac{\pi}{4}\varphi(x)}(-1)^{f_3(x)}\ket{x}$. Toffoli-count optimization targets $f_3$, while $\varphi$ is retained for subsequent CNOT-count optimization.

Minimizing the $T$-count is equivalent to finding a Waring decomposition of the signature tensor $S \in \F_2^{n \times n \times n}$ with the fewest terms~\cite{amy_2019,heyfron_2018}. Each term corresponds to one $T$ gate acting on a parity of input qubits (Fig.~\ref{fig:workflow}b). Minimizing the Toffoli count is equivalent to finding a canonical polyadic (CP) decomposition of the cubic tensor $T$:
\begin{equation}
    T_{ijk} = \sum_{q=1}^{\sub{r}{cp}} U_{qi}\, V_{qj}\, W_{qk},
    \label{eq:cp-decomp}
\end{equation}
where each rank-1 term $(U_q, V_q, W_q)$ encodes the three parities for one $CCZ$ gate (Fig.~\ref{fig:workflow}c). The number of terms $\sub{r}{cp}$ equals the Toffoli count of the corresponding circuit. The CP rank of $T$ is the minimum value of $\sub{r}{cp}$ over all its CP decompositions; unless stated otherwise, reported rank values denote the number of terms in the corresponding decomposition, with no claim of optimality.

\begin{figure*}
    \centering
    \includegraphics{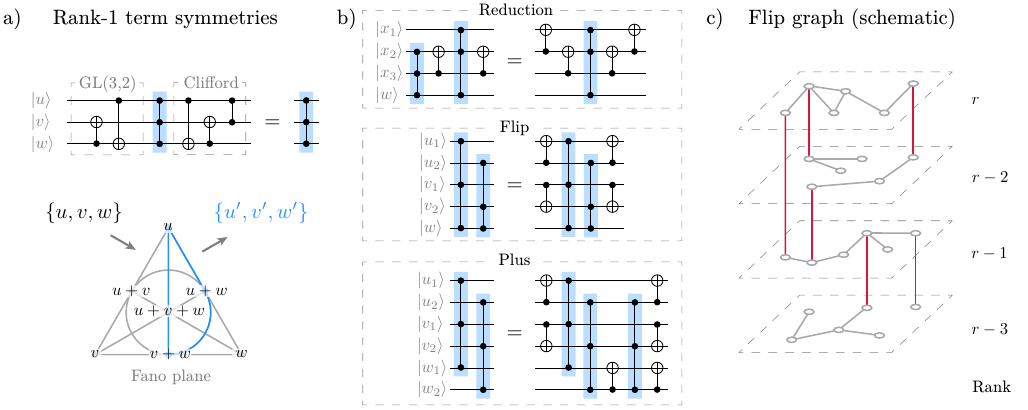}
\caption{\justifying
\textbf{Flip graph symmetries, operations, and structure.}
a) Local CNOT basis changes before a $CCZ$ gate can be compensated by Clifford gates. Algebraically, a cubic rank-1 term $(u,v,w)$ depends only on the three-dimensional subspace $\langle u,v,w\rangle$: the invertible binary basis changes $\mathrm{GL}(3,2)$ preserve its cubic part. The Fano plane shows the seven nonzero combinations; any noncollinear triple, such as the blue triangle, represents the same cubic term.
b) The flip graph is generated by three local transformations. A \emph{reduction} decreases rank by collecting all terms over a shared factor and minimizing the induced quadratic form. A \emph{flip} rewires two terms sharing a common factor $w$ while preserving rank: $u_1 v_1 w + u_2 v_2 w \mapsto (u_1 + u_2) v_1 w + u_2 (v_1 + v_2) w$. A \emph{plus} combines an inverse reduction with a flip to escape local plateaus, temporarily increasing the rank: $u_1 v_1 w_1 + u_2 v_2 w_2 \mapsto (u_1 + u_2) v_1 w_1 + u_2 (v_1 + v_2) w_2 + u_2 v_1 (w_1 + w_2)$.
c) Schematic flip graph structure: vertices are schemes grouped by rank (number of terms) on dashed planes; horizontal grey edges correspond to flips, and vertical red edges to reductions.
Blue shading in (a,b) marks $CCZ$ gates.
}
    \label{fig:flip-graph}
\end{figure*}

Over $\F_2$, the cubic contribution of each rank-1 term admits a $\mathrm{GL}(3,2)$ symmetry: any invertible transformation of the triple $(U_q, V_q, W_q)$ preserves its cubic part, while the induced lower-degree phase is absorbed into~$\varphi$. In circuit terms, local CNOTs applied before a $CCZ$ gate can always be compensated by Clifford gates (Fig.~\ref{fig:flip-graph}a). The non-Clifford content of $T$ is thus captured by its alternating part $\mathrm{Alt}(T)$, making Toffoli minimization formally a skew-symmetric tensor decomposition problem~\cite{arrondo_2021}.

The two decomposition problems are related: each $CCZ$ expands into $7$ $T$ gates, giving $\sub{r}{w} \leq 7\,\sub{r}{cp}$, though the expanded representation typically admits further rank reduction. Toffoli-count optimization for reversible circuits and $T$-count optimization for Clifford+$T$ circuits are NP-hard in general~\cite{vandewetering_2024,hillar_2013}, motivating the heuristic methods developed below.

\subsection{Toffoli-count minimization} \label{sec:results:b}

We develop a three-stage pipeline for Toffoli-count minimization: basis change optimization (BCO), symplectic Gaussian elimination (SGE), and flip graph search (FGS). Each stage operates on successively richer search spaces.

BCO searches over linear changes of basis applied before the phase operator, corresponding to a circuit architecture where all CNOTs precede all $CCZ$ gates (Fig.~\ref{fig:workflow}a). The goal is to find a basis in which the cubic polynomial $f_3$ has the fewest monomials; this minimum is the algebraic thickness~\cite{carlet_2002}. The search uses greedy or beam search over elementary substitutions
\begin{equation}
    x_i \leftarrow x_i \oplus x_j,
\end{equation}
each corresponding to a single CNOT gate. For most benchmark circuits from~\cite{ruiz_2025}, BCO alone suffices to match or improve prior results.

SGE exploits the $\mathrm{GL}(3,2)$ symmetry of rank-1 terms to reduce the Toffoli count further. When multiple $CCZ$ terms share a common element in their subspaces, their combined contribution factorizes into a linear form times a quadratic form. Minimizing the $CCZ$ count then reduces to minimizing the rank of this quadratic form, which can be computed via symplectic Gaussian elimination in $O(n^3)$ time~\cite{witt_1937}. Applied iteratively, SGE greedily selects the shared factor with the largest rank reduction. For all benchmark cubic circuits in Table~\ref{tab:a-bench}, BCO followed by SGE suffices to match or improve prior results.

Algebraic thickness and CP rank are not equal in general. A minimal example is
\begin{equation}
    x_1 x_3 x_5 + (x_1 + x_2) (x_3 + x_4) x_6 + x_2 x_4 (x_5 + x_6),
\end{equation}
which has algebraic thickness~5 but CP rank~3~\cite{hora_2021}. To find decompositions below algebraic thickness and SGE, we introduce flip graph search (FGS), adapting the framework developed for matrix multiplication~\cite{kauers_2023,moosbauer_2025} to cubic phase polynomials. The flip graph at rank~$r$ has vertices corresponding to CP decompositions with $r$ terms, connected by three local transformations (Fig.~\ref{fig:flip-graph}b): \emph{flips} rewire two terms sharing a common factor while preserving rank; \emph{reductions} decrease rank via SGE on shared factors; and \emph{plus} transitions~\cite{kauers_2023} temporarily increase rank to escape local plateaus. The search proceeds via random walks on this graph, maintaining a diverse pool of decompositions at each rank level (Fig.~\ref{fig:flip-graph}c).

\begin{figure*}
    \centering
    \includegraphics{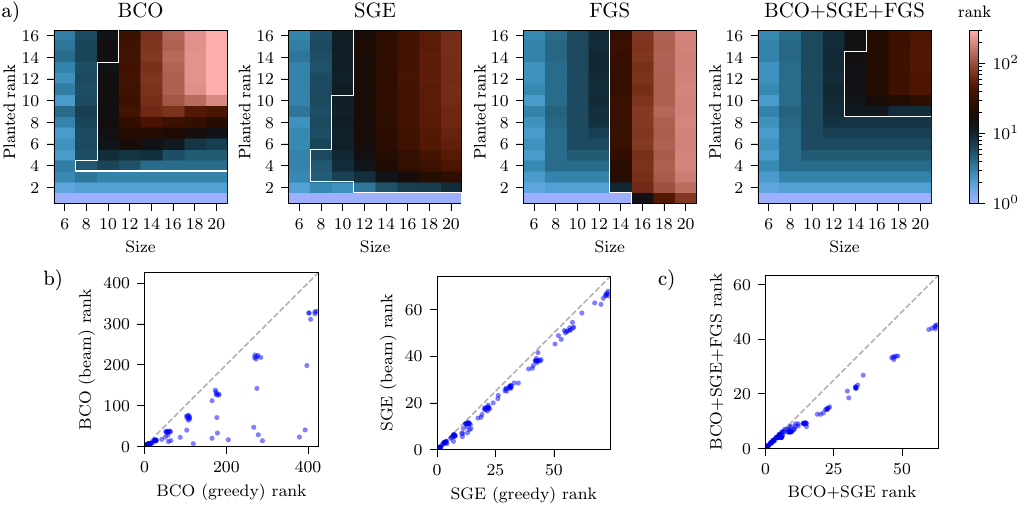}
\caption{\justifying
\textbf{Performance on the planted CP benchmark.}
For each tensor dimension $n$ (Size) and planted rank $\sub{r}{pl}$, we form a canonical polyadic (CP) decomposition from uniformly sampled binary factor matrices $U,V,W$ and form $T_{ijk}=\sum_{q=1}^{\sub{r}{pl}}U_{qi}V_{qj}W_{qk}$.
We then run the decomposition methods on $T$ and report the recovered rank, averaged over 4 independently generated tensors for each parameter pair.
a) Heatmaps show the recovered rank for basis change optimization (BCO), symplectic Gaussian elimination (SGE), flip graph search (FGS), and their successive application BCO+SGE+FGS, using a shared logarithmic color scale. The white contour separates the region where the recovered rank exceeds the planted upper bound by more than one.
b) Greedy vs.\ beam-search variants of BCO and SGE, using beam width $2^{10}$.
c) Rank improvement from FGS after BCO+SGE.
Dashed lines indicate equality.
}
    \label{fig:rnd-bench}
\end{figure*}

Fig.~\ref{fig:rnd-bench} shows evaluation on planted CP benchmarks with known structure. For each dimension~$n$ and planted rank~$\sub{r}{pl}$, we sample random tensors and attempt to recover the planted rank. BCO alone recovers low ranks for small tensors but plateaus as dimension and rank increase; adding SGE extends the region of successful recovery, and FGS further expands it (Fig.~\ref{fig:rnd-bench}a). Beam search consistently outperforms greedy selection for both BCO and SGE (Fig.~\ref{fig:rnd-bench}b).  The computational cost of each stage differs: producing the planted benchmark results for all 512 tensors in Fig.~\ref{fig:rnd-bench} required a cumulative wall-clock time equivalent to 23 minutes for BCO, 9 minutes for SGE, and approximately 9 days for FGS on a single 48-thread CPU. 

For circuits implementing bilinear maps, such as finite field
multiplication, the phase polynomial has a special structure:
the cubic tensor decomposes as a product over disjoint input
and output registers, yielding a smaller tensor of dimensions
$(n_a, n_b, n_c)$ rather than $(n, n, n)$ for $n = n_a + n_b + n_c$
(Fig.~\ref{fig:gf-mult}).
Flip graph search can operate directly on this smaller tensor,
exponentially reducing the search space.
This connects to recent work on polynomial multiplication
over $\F_2$~\cite{chen_2025} and improves Toffoli counts
for $\mathrm{GF}(2^p)$ multiplication circuits.
We optimize both the convolution tensor and the field tensor
and report the best result.

Table~\ref{tab:a-bench} presents results on 26 benchmark circuits from~\cite{ruiz_2025}. BCO+SGE with beam search improves upon the best prior Toffoli counts in 7 of the 25 cubic circuits, with the remaining circuits matching prior results. FGS finds no further improvement on these circuits. Our approach optimizes the phase polynomial directly without circuit splitting, yielding implementations whose non-Clifford gates are all Toffoli, whereas~\cite{ruiz_2025} sometimes required mixed gate sets with residual $T$ and $CS$ gates.

Table~\ref{tab:gf-mult} presents results for $\mathrm{GF}(2^p)$
multiplication circuits. Multiplication over $\mathrm{GF}(2^p)$ has multiplicative complexity linear in $p$, with the uniform bound $\mu_2(p) \leq \frac{189}{22}p + 18$~\cite{pieltant_2015} and asymptotic bound~7~\cite{rambaud_2017,ballet_2021}. Such decompositions map to quantum circuits with the same Toffoli count~\cite{liao_2025}. Our bilinear method does not exploit the recursive structure behind these bounds, so these instances test how closely a general automatic search can approach known constructions. The general pipeline (SGE+FGS on the full tensor) matches prior results for small fields but struggles for $p \geq 7$. The bilinear formulation succeeds where the general approach does not, matching or improving all prior automated results~\cite{ruiz_2025} and reproducing most bounds of~\cite{chen_2025}; searching the field tensor directly can yield lower ranks than the convolution tensor alone, e.g.\ rank~15 versus~17 for $p=6$. The degree-specific bounds reported in~\cite{cenk_2010} for $p \leq 10$ have not been improved since 2009. Our automatic search matches them through $p \leq 7$, but remains above them for $p=8,9,10$.

An additional benchmark suite of 35~circuits from~\cite{ruiz_2025},  covering binary addition, Hamming-weight phasing, unary iteration,  and quantum chemistry, spanning sizes up to 72~qubits, is matched by BCO+SGE within one minute per circuit. FGS finds no further improvement.

\begin{table*}
    \renewcommand{\arraystretch}{0.94}
    \caption{\justifying
    \textbf{Benchmark results for non-Clifford gate minimization.}
    Comparison of Toffoli-count ($r_{\mathrm{cp}}$) and $T$-count ($r_w$) optimization on circuits from~\cite{ruiz_2025}.
    Here $n$ is the original qubit count, $h$ is the number of internal Hadamard gates after preprocessing, and Size is the phase-block qubit count $n+h$. The Toffoli-optimization runtime $t$ is reported in seconds.
    The QFT$_4$ construction is described in the Supplementary Information.
    For $T$ minimization, CP decompositions generated by flip graph search initialize our implementation of the optimizer in~\cite{heyfron_2018,vandaele_2025a}.
    Bold entries indicate improvements over the best prior results. The grey $+T$ and $+CS$ columns report additional $T$ and controlled-$S$ gates. Blank gate-count entries denote zero.
    BCO denotes basis change optimization, and SGE symplectic Gaussian elimination.
    $^\dag$Result obtained using BCO+SGE; unmarked entries use BCO only.
    }
    \begin{tabular}{lrrrrrrrrrrrrrrr}
            \toprule
            & & & &
            \phantom{4} &
            \multicolumn{7}{c}{Toffoli-minimization ($r_\mathrm{cp}$)}
            &
            \phantom{4} &
            \multicolumn{3}{c}{$T$-minimization ($r_\mathrm{w}$)}
            \\
            \cmidrule(lr){6-12} \cmidrule(lr){14-16}
            &&&&&
            \multicolumn{3}{c}{Best prior\textsuperscript{\cite{ruiz_2025}}}
            &
            \multicolumn{4}{c}{This work}
            &&
            Best prior\textsuperscript{[\citenum{ruiz_2025},\citenum{vandaele_2025a}]}
            &
            \multicolumn{2}{c}{This work}
            \\
            \cmidrule(lr){6-8} \cmidrule(lr){9-12} \cmidrule(lr){14-14} \cmidrule(lr){15-16}
            Circuit &
            \grey{$n$} &
            \grey{$h$} &
            Size &
            &
            {\footnotesize \grey{$+T$}} &
            {\footnotesize \grey{$+CS$}} &
            Count &
            {\footnotesize \grey{$+T$}} &
            {\footnotesize \grey{$+CS$}} &
            Count\tdago &
            \grey{$t$,\,s} &
            &
            Count\phantom{\tcc} &
            Count &
            \\
            \midrule
Adder${}_8$                & \grey{24} & \grey{ 37} &  61 & & \grey{ 28} & \grey{  } &   33 & \grey{  } & \grey{  } &   \textbf{27}\tdagl &  \grey{ 68}  & & 119\ttcb & \textbf{117} &   \\
Barenco Tof${}_3$          & \grey{ 5} & \grey{  3} &  8  & & \grey{   } & \grey{  } &    2 & \grey{  } & \grey{  } &            2\tdago  &  \grey{0}    & &  13\ttcc &         13   &   \\
Barenco Tof${}_4$          & \grey{ 7} & \grey{  7} &  14 & & \grey{   } & \grey{  } &    4 & \grey{  } & \grey{  } &            4\tdago  &  \grey{1}    & &  23\ttcc &         23   &   \\
Barenco Tof${}_5$          & \grey{ 9} & \grey{ 11} &  20 & & \grey{   } & \grey{  } &    6 & \grey{  } & \grey{  } &            6\tdago  &  \grey{1}    & &  33\ttcc &         33   &   \\
Barenco Tof${}_{10}$       & \grey{19} & \grey{ 31} &  50 & & \grey{   } & \grey{  } &   16 & \grey{  } & \grey{  } &           16\tdago  &  \grey{12}   & &  83\ttcc &         83   &   \\
CSLA MUX${}_3$             & \grey{15} & \grey{  6} &  21 & & \grey{   } & \grey{  } &    8 & \grey{  } & \grey{  } &            8\tdago  &  \grey{1}    & &  39\ttcc &         39   &   \\
CSUM MUX${}_9$             & \grey{30} & \grey{ 12} &  42 & & \grey{   } & \grey{  } &   14 & \grey{  } & \grey{  } &           14\tdago  &  \grey{6}    & &  71\ttcc &         71   &   \\
Grover${}_5$               & \grey{ 9} & \grey{ 68} &  77 & & \grey{ 12} & \grey{  } &   27 & \grey{  } & \grey{  } &   \textbf{25}\tdago &  \grey{101}  & & 143\ttcb &         143  &   \\
Ham${}_{15}$ (low)         & \grey{17} & \grey{ 25} &  42 & & \grey{   } & \grey{  } &   17 & \grey{  } & \grey{  } &           17\tdago  &  \grey{ 16}  & &  73\ttca &         73   &   \\
Ham${}_{15}$ (med)         & \grey{17} & \grey{ 54} &  71 & & \grey{  8} & \grey{  } &   35 & \grey{  } & \grey{  } &   \textbf{33}\tdagl &  \grey{ 24}  & & 137\ttcb &         137  &   \\
Ham${}_{15}$ (high)        & \grey{20} & \grey{331} & 351 & & \grey{ 90} & \grey{ 2} &  173 & \grey{  } & \grey{  } &  \textbf{155}\tdagl &  \grey{3430} & & 643\ttcb &          643 &   \\
HWB${}_6$                  & \grey{ 7} & \grey{ 20} &  27 & & \grey{   } & \grey{  } &   10 & \grey{  } & \grey{  } &           10\tdagl  &  \grey{3}    & &  51\ttcc &         51   &   \\
Mod 5${}_4$                & \grey{ 5} & \grey{  0} &  5  & & \grey{   } & \grey{  } &    1 & \grey{  } & \grey{  } &            1\tdago  &  \grey{0}    & &   7\ttcc &         7    &   \\
Mod Adder${}_{1024}$       & \grey{28} & \grey{304} & 332 & & \grey{188} & \grey{15} &  141 & \grey{  } & \grey{  } &  \textbf{128}\tdago &  \grey{15800}& & 575\ttcb & \textbf{573} &   \\
Mod Mult${}_{5}$           & \grey{ 9} & \grey{  2} &  11 & & \grey{   } & \grey{  } &    3 & \grey{  } & \grey{  } &            3\tdago  &  \grey{0}    & &  17\ttcc &         17   &   \\
Mod Red${}_{21}$           & \grey{11} & \grey{ 17} &  28 & & \grey{   } & \grey{  } &   11 & \grey{  } & \grey{  } &           11\tdago  &  \grey{4}    & &  51\ttcc &         51   &   \\
QCLA Adder${}_{10}$        & \grey{36} & \grey{ 25} &  61 & & \grey{ 28} & \grey{ 5} &   28 & \grey{  } & \grey{  } &   \textbf{24}\tdago &  \grey{5}    & & 109\ttcb & \textbf{107} &   \\
QCLA Com${}_{7}$           & \grey{24} & \grey{ 18} &  42 & & \grey{   } & \grey{  } &   12 & \grey{  } & \grey{  } &           12\tdago  &  \grey{11}   & &  59\ttcc &         59   &   \\
QCLA Mod${}_{7}$           & \grey{26} & \grey{ 58} &  84 & & \grey{ 36} & \grey{  } &   43 & \grey{  } & \grey{  } &   \textbf{37}\tdago &  \grey{ 55}  & & 159\ttcb & \textbf{153} &   \\
QFT${}_{4}$                & \grey{ 5} & \grey{ 38} &  43 & & \grey{ 30} & \grey{ 3} &    4 & \grey{30} & \grey{ 3} &            4\tdago  &  \grey{10}   & &  53\ttcc &         53   &   \\
RC Adder${}_{6}$           & \grey{14} & \grey{ 10} &  24 & & \grey{   } & \grey{  } &    6 & \grey{  } & \grey{  } &            6\tdago  &  \grey{2}    & &  37\ttcc &         37   &   \\
Tof${}_{3}$                & \grey{ 5} & \grey{  2} &  7  & & \grey{   } & \grey{  } &    2 & \grey{  } & \grey{  } &            2\tdago  &  \grey{0}    & &  13\ttcc &         13   &   \\
Tof${}_{4}$                & \grey{ 7} & \grey{  4} &  11 & & \grey{   } & \grey{  } &    3 & \grey{  } & \grey{  } &            3\tdago  &  \grey{1}    & &  19\ttcc &         19   &   \\
Tof${}_{5}$                & \grey{ 9} & \grey{  6} &  15 & & \grey{   } & \grey{  } &    4 & \grey{  } & \grey{  } &            4\tdago  &  \grey{1}    & &  25\ttcc &         25   &   \\
Tof${}_{10}$               & \grey{19} & \grey{ 16} &  35 & & \grey{   } & \grey{  } &    9 & \grey{  } & \grey{  } &            9\tdago  &  \grey{7}    & &  55\ttcc &         55   &   \\
VBE Adder${}_{3}$          & \grey{10} & \grey{  4} &  14 & & \grey{   } & \grey{  } &    3 & \grey{  } & \grey{  } &            3\tdago  &  \grey{1}    & &  19\ttcc &         19   &   \\
        \bottomrule
    \end{tabular}
    \label{tab:a-bench}

\caption{\justifying
\textbf{Benchmark results for GF$(2^p)$ multiplication circuits.}
For prior work, Auto denotes automatic optimization; Overall reports the best known results, including the constructions of~\cite{cenk_2010,chen_2025} for Toffoli minimization.
For Toffoli minimization, General applies symplectic Gaussian elimination and flip graph search to the full phase tensor, while Bilinear works with smaller field-multiplication or convolution tensors.
For $T$ minimization, the optimizer~\cite{heyfron_2018,vandaele_2025a} is initialized from canonical polyadic decompositions: Auto uses decompositions obtained by the General and Bilinear approaches, while +MC uses known multiplicative-complexity constructions from~\cite{cenk_2010}.
Size is the qubit count $3p$; $r_{\mathrm{cp}}$ and $r_w$ denote the reported Toffoli and $T$ counts.
Bold entries improve upon the best prior results.
$^*$Result not achieved in every run (at least 1 of 10).
$^\dag$Requires Bilinear initialization.
$^\ddag$Requires basis change optimization preprocessing and the General approach.
}
    \begin{tabular}{lrrrrrrrrrrr}
        \toprule
        &
        &
        &
        \multicolumn{4}{c}{Toffoli minimization ($r_\mathrm{cp}$)}
        &
        \multicolumn{5}{c}{$T$ minimization ($r_\mathrm{w}$)}
        \\
        \cmidrule(lr){4-7} \cmidrule(lr){8-12}
        &
        &
        &
        \multicolumn{2}{c}{Best prior}
        &
        \multicolumn{2}{c}{This work}
        &
        \multicolumn{3}{c}{Best prior}
        &
        \multicolumn{2}{c}{This work}
        \\
        \cmidrule(lr){4-5} \cmidrule(lr){6-7} \cmidrule(lr){8-10} \cmidrule(lr){11-12}
        Circuit &
        Size &
        &
        \grey{Auto\textsuperscript{\cite{ruiz_2025}}} &
        Overall\phantom{\textsuperscript{[\citenum{ruiz_2025},\citenum{vandaele_2025a},\cite{cenk_2010}]}} &
        General\phantom{$^*$} &
        Bilinear\phantom{$^*$} &

        \grey{Auto\textsuperscript{[\citenum{ruiz_2025}]}} &
        \grey{Auto\textsuperscript{[\citenum{vandaele_2025a}]}} &
        Overall\textsuperscript{[\citenum{ruiz_2025},\citenum{vandaele_2025a}]} &
        Auto\tstro\tdago\tddgo &
        +MC\textsuperscript{[\citenum{cenk_2010}]}
        \\
        \midrule
GF$(2^2 )$ Mult   &   6 & &  \grey{ 3}\tatqo &   3\tcc &   3\phantom{$^*$} &           3\phantom{$^*$} & \grey{ 17}\tatqo & \grey{ 18}\tvvvo &  17\ttcc &           17\tstro\tdago\tddgo &  17\tcf \\
GF$(2^3 )$ Mult   &   9 & &  \grey{ 6}\tatqo &   6\tcc &   6\phantom{$^*$} &           6\phantom{$^*$} & \grey{ 29}\tatqo & \grey{ 36}\tvvvo &  23\ttcb &           29\tstro\tdago\tddgo &  29\tcf \\
GF$(2^4 )$ Mult   &  12 & &  \grey{ 9}\tatqo &   9\tcc &   9\phantom{$^*$} &           9\phantom{$^*$} & \grey{ 39}\tatqo & \grey{ 49}\tvvvo &  39\ttca &           39\tstrl\tdago\tddgo &  45\tcf \\
GF$(2^5 )$ Mult   &  15 & &  \grey{13}\tatqo &  13\tca &  13\phantom{$^*$} &          13\phantom{$^*$} & \grey{ 59}\tatqo & \grey{ 81}\tvvvo &  59\ttca &           59\tstrl\tddgl\tdago &  63\tcf \\
GF$(2^6 )$ Mult   &  18 & &  \grey{18}\tatqo &  15\tce &  15$^*$  & 15\phantom{$^*$} & \grey{ 77}\tatqo & \grey{113}\tvvvo &  77\ttca &           77\tdagl\tstro\tddgo &  81\tcf \\
GF$(2^7 )$ Mult   &  21 & &  \grey{22}\tatqo &  22\tca &  23$^*$           &          22\phantom{$^*$} & \grey{104}\tatqo & \grey{155}\tvvvo & 104\ttca & \textbf{101}\tdagl\tstro\tddgo &  103\tcf \\
GF$(2^8 )$ Mult   &  24 & &  \grey{29}\tatqo &  24\tce &  34$^*$           & 26\phantom{$^*$} & \grey{123}\tatqo & \grey{205}\tvvvo & 123\ttca &          123\tdagl\tstrl\tddgo &  127\tcf \\
GF$(2^9 )$ Mult   &  27 & &  \grey{35}\tatqo &  30\tce &  51$^*$           & 32$^*$           & \grey{161}\tatqo & \grey{257}\tvvvo & 161\ttca & 153\tdagl\tstrl\tddgo &  \textbf{147}\tcf \\
GF$(2^{10})$ Mult &  30 & &  \grey{46}\tatqo &  33\tce &  66$^*$           & 39\phantom{$^*$} & \grey{196}\tatqo & \grey{315}\tvvvo & 196\ttca & 185\tdagl\tstro\tddgo &  \textbf{173}\tcf \\
        \bottomrule
    \end{tabular}
    \label{tab:gf-mult}
\end{table*}

\subsection{CNOT-count optimization} \label{sec:results:cnot}

Across the 25 cubic benchmark circuits in Table~\ref{tab:a-bench}, the AlphaTensor-Quantum baseline~\cite{ruiz_2025} has a total phase-block CNOT count of $52\,391$. For comparison, realizing each term of our decompositions independently by compute--$CCZ$--uncompute and implementing the remaining quadratic phase directly uses $27\,296$ CNOT gates. Our decompositions contain 560 Toffoli gates in total, and the CPR routes use $13\,432$ CNOT gates, including basis restoration. Implementing the remaining quadratic phase directly in the input coordinates adds $6\,543$ CNOT gates. QPR reduces this contribution to $3\,824$, giving a total CNOT count of $17\,256$. Per-circuit results are given in Supplementary Table~1.

For bilinear maps $c\mapsto c+g(a,b)$, CPR routes parities separately within the three registers, reducing the search space. Applying CPR to $\mathrm{GF}(2^p)$ multiplication decompositions based on~\cite{cenk_2010} gives the circuits in Table~\ref{tab:gf-mult-cnot-short}. For $p=2,4,8$, our circuits match or improve both gate counts reported in~\cite{yosri_2026}. At $p=4$, our circuit uses 9 Toffoli and 26 CNOT gates, compared with 9 Toffoli and 27 CNOT gates in~\cite{luo_2025} and 9 Toffoli and 41 CNOT gates in~\cite{yosri_2026}. At $p=8$, our circuit uses 24 Toffoli and 137 CNOT gates, compared with 27 Toffoli and 118 CNOT gates in~\cite{luo_2025} for a zero-initialized output. At $p=12$, our circuit uses 42 Toffoli and 308 CNOT gates, compared with 51 Toffoli and $1\,506$ CNOT gates reported in~\cite{khattar_2025} for a zero-initialized output. For $p=16,32,64,128$, our circuits reduce the Toffoli count relative to~\cite{yosri_2026}, but use more CNOT gates. At $p=16$, the Toffoli count decreases from 81 to 60, while the CNOT count increases from 615 to 690.

\begin{figure*}
    \centering
    \includegraphics[width=\textwidth]{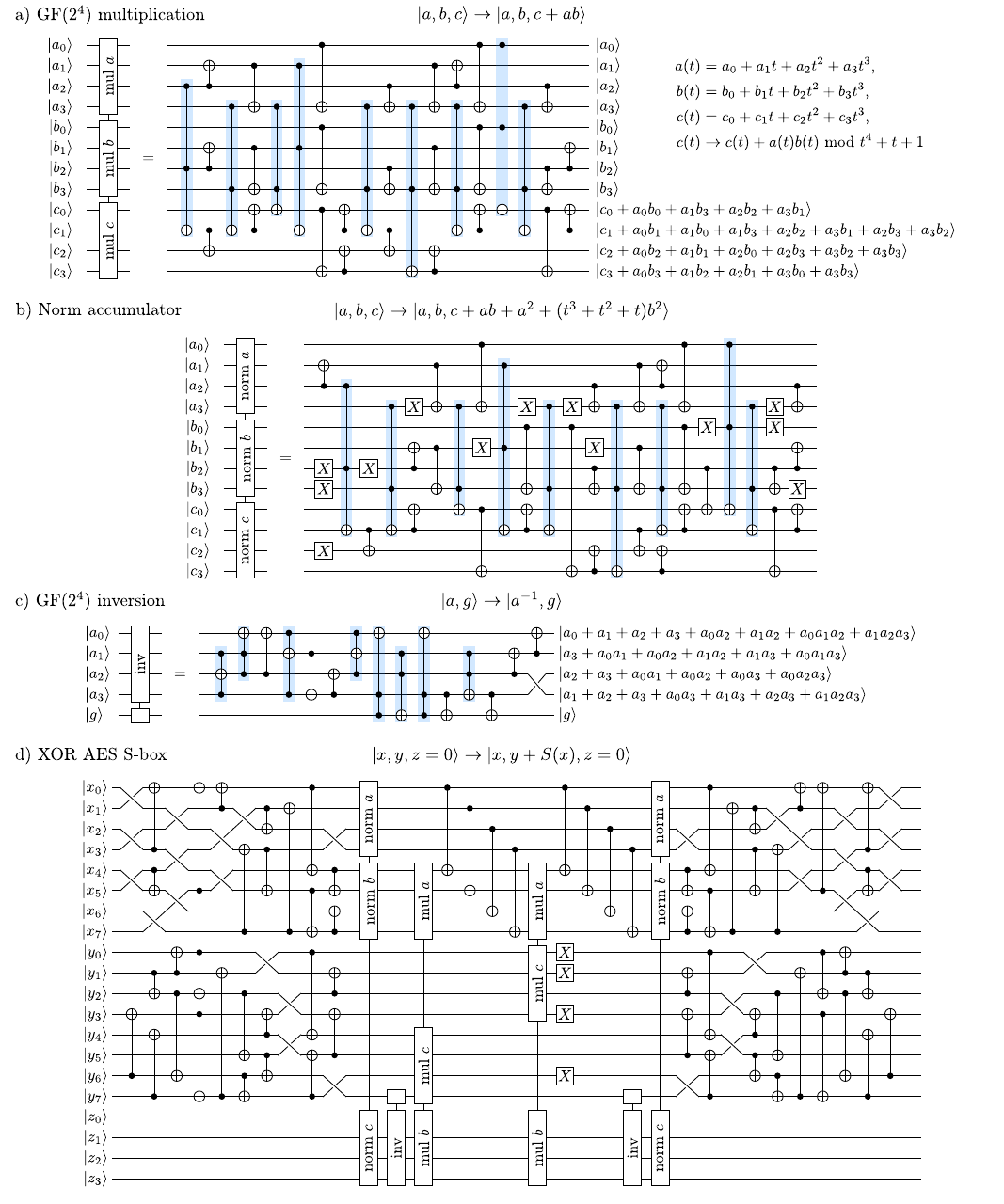}
\caption{\justifying
\textbf{S-box quantum circuit.}
a) A multiplication accumulator over the finite field $\mathrm{GF}(2^4)$ (mult), with 9 Toffoli and 26 CNOT gates.
b) A norm accumulator with 9 Toffoli and 28 CNOT gates, obtained from (a) by adding linear terms.
c) $\mathrm{GF}(2^4)$ inversion (inv), with $0^{-1}=0$, preserving the borrowed qubit $g$ and using 8 Toffoli and 7 CNOT gates~\cite{chen_2025b}.
d) The 20-qubit implementation of the Advanced Encryption Standard (AES) S-box $S$, with 52 Toffoli and 186 CNOT gates, assembled from two copies of each block in (a)--(c) and affine transformations. It implements $y\mapsto y+S(x)$ by bitwise XOR; $x,y$ are eight-qubit registers, and the four workspace qubits $z$ are restored to zero.
Blue shading marks Toffoli gates.
}
    \label{fig:aes-sbox-building-blocks}
\end{figure*}

The 20-qubit XOR AES S-box circuits in~\cite{li_2023,chen_2025b} compute inversion in $\mathrm{GF}(2^8)$ using arithmetic over $\mathrm{GF}(2^4)$. These constructions use a multiplication accumulator and a norm accumulator obtained by adding linear terms. Applying CPR and QPR gives these blocks with 26 and 28 CNOT gates, respectively (Fig.~\ref{fig:aes-sbox-building-blocks}a,b), compared with 28 and 41 CNOT gates in~\cite{li_2023}. The resulting block-structured circuit in Fig.~\ref{fig:aes-sbox-building-blocks}d uses 52 Toffoli and 186 CNOT gates. Starting from this circuit, we obtain two manually optimized variants: one with 147 CNOT gates and another with Toffoli depth 24 and 184 CNOT gates. Table~\ref{tab:aes-sbox-cnot} compares the three circuits with prior 20-qubit constructions.

\begin{figure*}
    \centering
    \includegraphics{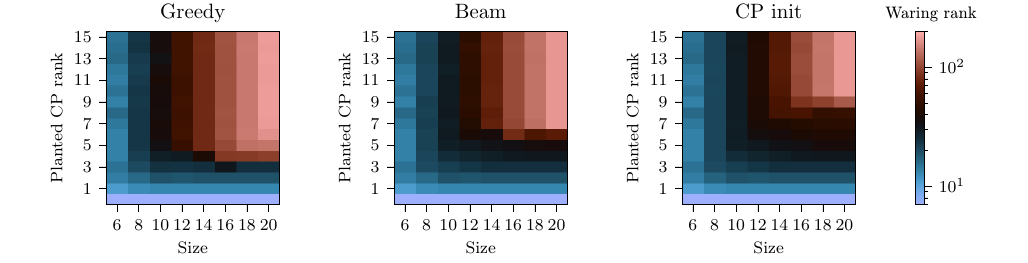}
\caption{\justifying
\textbf{Waring decomposition performance on the planted CP benchmark.}
For the same planted canonical polyadic (CP) tensors as in Fig.~\ref{fig:rnd-bench}, we seek low-rank Waring decompositions of the associated signature tensors.
We compare three approaches, from left to right: greedy selection in FastTODD~\cite{vandaele_2025a}, beam search with width $2^{10}$, and greedy selection with CP initialization (CP init).
The shared logarithmic color scale shows the recovered Waring rank $\sub{r}{w}$, averaged over 4 independently generated tensors for each parameter pair.
Size is the tensor dimension, and Planted CP rank is the number of terms in the planted decomposition.
}
    \label{fig:rnd-bench-waring}
\end{figure*}

\subsection{\texorpdfstring{$\boldsymbol{T}$}{T}-count minimization} \label{sec:results:c}

Decompositions obtained by Toffoli-count minimization provide effective initializations for $T$-count optimization via TODD~\cite{heyfron_2018} with the improvements of~\cite{vandaele_2025a}.

Each CP term $(u, v, w)$ expands into seven Waring terms $\{u,\, v,\, w,\, u \oplus v,\, u \oplus w,\, v \oplus w,\, u \oplus v \oplus w\}$, giving the naive bound $\sub{r}{w} \leq 7\,\sub{r}{cp}$. The expanded representation typically admits substantial rank reduction: TODD iteratively applies rank-reducing transformations $A \leftarrow A \oplus y z\T$ to the gate synthesis matrix, removing duplicate or zero rows until no further reduction is possible. Since TODD is greedy, its outcome depends strongly on initialization. The flip graph search from the Toffoli pipeline typically produces around $10^3$ distinct CP decompositions; expanding them provides diverse starting points that guide TODD into different regions of the search space.

Fig.~\ref{fig:rnd-bench-waring} evaluates this strategy on planted instances. CP initialization consistently outperforms both naive greedy and beam search with width $2^{10}$. Notably, combining beam search with CP initialization yields no further improvement: diverse starting points from the flip graph appear to compensate for the lack of within-run exploration.

On benchmark circuits (Table~\ref{tab:a-bench}), CP-initialized TODD improves the $T$-count in 4 cases compared to the best prior results from~\cite{ruiz_2025,vandaele_2025a}, with the remaining circuits matching prior performance. To maximize diversity of initializations, the CP decompositions here were generated by BCO+FGS alone, bypassing SGE; as a result, the $T$-count results in Table~\ref{tab:a-bench} required a cumulative wall-clock time equivalent to approximately 54 hours on a single 48-thread CPU.

For $\mathrm{GF}(2^p)$ multiplication (Table~\ref{tab:gf-mult}), using known multiplicative complexity constructions from~\cite{cenk_2010} as CP initialization for TODD improves $T$-counts for $p = 9$ and $p = 10$.

\section{Discussion} \label{sec:discussion}

Our approach prioritizes Toffoli count before optimizing CNOT count, leaving the Toffoli--CNOT Pareto frontier largely unexplored. Decompositions of the same rank can have different CNOT counts, and a higher-rank decomposition may require fewer CNOTs. CPR and QPR can be applied to intermediate decompositions produced by FGS, not only its final output. Greedy CPR could provide an inexpensive estimate for selecting candidates by CNOT count during FGS. We leave a systematic study of these trade-offs to future work.

The standard cubic benchmark collection is relatively small and heterogeneous, and BCO+SGE already matches or improves its best reported Toffoli counts. We used fixed search parameters for BCO, SGE, and FGS, without per-instance tuning. The difficulty of finite field multiplication suggests that more demanding, systematic benchmarks are needed. Our planted tensors are a step in this direction, allowing dimension and planted rank to be varied independently. Extending them beyond uniformly sampled factors, and adding scalable families from quantum algorithms, would help assess which structures make the search difficult.

An important next step for $\mathrm{GF}(2^p)$ multiplication is to combine linear Toffoli count with quasilinear total gate count. Fast in-place algorithms for polynomial accumulation and modular reduction~\cite{dumas_2026} suggest a route toward this goal.

In~\cite{khattar_2025}, PCTOF gates serve as primitives for lattice surgery, and their number is reduced using linear dependencies among products of input parities. A dependence among $U_q\otimes V_q$ allows a term to be removed (and analogously for the other two factor pairs). Our bilinear search restricts reduction checks to groups sharing a factor to keep them inexpensive during the random walks. Applying broader dependency checks may be useful, particularly for larger bilinear circuits.

The manually optimized AES S-box variants show the benefit of combining linear transformations across block boundaries and rearranging gates. Automating these steps, including optimization of non-Clifford depth~\cite{selinger_2013}, is a natural direction for future work.

Finally, the flip graph and the space of basis changes provide structured settings for learned heuristics~\cite{chervov_2025}. Neural guidance could select among algebraically valid transformations, retaining correctness while learning which paths are more likely to reduce circuit cost.

\section{Methods} \label{sec:methods}

\subsection{Phase polynomial formalism} \label{sec:methods:a}

The Hadamard gadgetization procedure~\cite{heyfron_2018}
replaces each internal Hadamard gate with a gadget consuming
one ancilla qubit initialized in $\ket{+}$,
with a Clifford correction conditioned on a measurement outcome.
Since Clifford corrections commute past the non-Clifford portion
up to a change of basis, they can be deferred to the end of the circuit.
The resulting CNOT+$T$ subcircuit decomposes into a diagonal
phase operator $U_f$ and a linear transformation $U_E$~\cite{amy_2014}:
\begin{align}
    U_f \ket{x} &= e^{i\frac{\pi}{4} f(x)} \ket{x}, \label{eq:phase-op} \\
    U_E \ket{x} &= \ket{Ex}, \label{eq:linear-op}
\end{align}
where $f\colon \F_2^n \to \Z_8$ is the phase function
and $E \in \mathrm{GL}(n, 2)$ is an invertible binary matrix.

Each $T$ gate acts on a parity of the input qubits
determined by preceding CNOTs, giving
\begin{equation}
    f(x) = \sum_{q=1}^{\sub{r}{w}} \bigg(\bigoplus_{i=1}^{n} A_{qi} x_i\bigg),
    \label{eq:xor-sum}
\end{equation}
where $A \in \F_2^{\sub{r}{w} \times n}$ is the gate synthesis
matrix~\cite{heyfron_2018} with one row per $T$ gate,
and the outer sum is over integers.
Conversely, any matrix $A$ specifies a CNOT+$T$ circuit:
for each row, CNOTs compute the parity on a target qubit,
$T$ is applied, and the CNOTs are reversed.
Since the phase $e^{i\frac{\pi}{4} f(x)}$ is $2\pi$-periodic,
$f$ is defined modulo~8.

To expand~\eqref{eq:xor-sum} into a standard polynomial,
we apply the identity
\begin{equation}
    x_i \oplus x_j = x_i + x_j - 2\, x_i x_j
    \label{eq:xor-expand}
\end{equation}
recursively.
Each additional variable in a parity introduces a factor of~2,
so working modulo~8, terms of degree~4 or higher vanish
(carrying a factor of at least~$2^3$).
The result is the weighted polynomial
\begin{equation}
    f(x) = \sum_i L_i x_i + 2\!\sum_{ij} Q_{ij} x_i x_j
    + 4\!\sum_{ijk} T_{ijk} x_i x_j x_k,
\end{equation}
where $L_i\in\Z_8$, $Q_{ij}\in\Z_4$, and $T_{ijk}\in\F_2$. Reducing $L_i$ and $Q_{ij}$ modulo~2 gives the coefficients of $f_1$ and $f_2$ in~\eqref{eq:weighted-poly}, while $f_3$ has coefficients $T_{ijk}$.

The remaining even parts define the diagonal Clifford phase
\begin{equation}
    \varphi = 2\,\varphi_1 + 4\,\varphi_2,
\end{equation}
where $\varphi_1$ is a linear polynomial with coefficients in $\Z_4$, and $\varphi_2$ is homogeneous quadratic over $\F_2$. Their coefficients are obtained by subtracting the coefficient parities from $L_i$ and $Q_{ij}$ and dividing the remainders by~2. The first part is realized by single-qubit $S$ powers, including $Z=S^2$, and the second by $CZ$ gates.

The gate correspondence follows from~\eqref{eq:xor-expand}. Matching the quadratic term shows that $CS$ uses $3$~$T$ gates; expanding $x_i \oplus x_j \oplus x_k$ produces $4\,x_i x_j x_k$, yielding a realization of $CCZ$ with $7$~$T$ gates~\cite{gosset_2014,amy_2013}.

The minimum $T$-count equals the Waring rank of the signature
tensor $S \in \F_2^{n \times n \times n}$ defined by
\begin{equation}
    S_{ijk} = \sum_{q=1}^{\sub{r}{w}} A_{qi}\, A_{qj}\, A_{qk}.
    \label{eq:signature-tensor}
\end{equation}
The tensor $S$ is symmetric: its diagonal $S_{iii}$ encodes $f_1$, partially-diagonal entries $S_{iij}$ encode $f_2$, and off-diagonal entries $S_{ijk}$ encode~$f_3$~\cite{heyfron_2018}.

For circuits with $f_1 = 0$, $f_2 = 0$, the CCZ--Toffoli equivalence is given by
\begin{equation}
    \mathrm{Toffoli}_{ab \to c}
    = (I_a \otimes I_b \otimes H_c)\, CCZ_{abc}\, (I_a \otimes I_b \otimes H_c).
\end{equation}
This setting covers nearly all standard benchmarks~\cite{ruiz_2025,vandaele_2025a,heyfron_2018} and connects to multiplicative complexity minimization in XAG synthesis~\cite{meuli_2019,meuli_2022}. The phase polynomial framework works with $CCZ$ gates directly, bypassing the compute/uncompute structure and ancilla management inherent in AND-count optimization.

\subsection{Basis change optimization} \label{sec:methods:b}

An elementary substitution $x_i \leftarrow x_i \oplus x_j$ corresponds to a CNOT gate acting before the phase operator. Composing such operations generates $\mathrm{GL}(n, 2)$, so the search targets a basis in which the non-Clifford part of the phase polynomial has the fewest monomials. This minimum is the algebraic thickness~\cite{carlet_2002}, and the corresponding circuit architecture places all CNOTs before all non-Clifford gates (Fig.~\ref{fig:workflow}a).

Under an elementary substitution, the non-Clifford components $(f_1,\,f_2,\,f_3)$ transform in a coupled manner. By~\eqref{eq:xor-expand}, products involving the substituted variable acquire additional terms at higher degrees: monomials from $f_1$ generate contributions to $f_2$, and monomials from $f_2$ generate contributions to $f_3$. For circuits with $f_1 = 0$ and $f_2 = 0$, the optimization reduces to minimizing the number of cubic monomials $|f_3|$, where each monomial requires one $CCZ$ gate. Algebraic thickness provides an upper bound on the CP rank: separated layers are a special case of interleaved circuits.

Exhaustive search over $\mathrm{GL}(n, 2)$ is infeasible: $|\mathrm{GL}(n, 2)| = \prod_{k=0}^{n-1}(2^n - 2^k)$ exceeds $10^{12}$ already for $n = 8$. In the greedy approach, we iteratively select the substitution $x_i \leftarrow x_i \oplus x_j$ that maximally decreases the cost, terminating when no improvement is possible. Each step evaluates $n(n-1)$ candidates. In beam search, we maintain a pool of the top $k$ candidates ranked by cost, expand each by all $n(n-1)$ substitutions, and retain the best $k$ results for the next iteration. All beam search experiments use width $k = 2^{10}$.

\subsection{Symplectic Gaussian elimination} \label{sec:methods:c}

Each rank-1 term $(u, v, w)$ in a CP decomposition spans a 3-dimensional subspace $\langle u, v, w \rangle$ whose seven nonzero elements form the \emph{seven-set}:
\begin{equation}
    \mathcal{S} = \{ u,\, v,\, w,\, u+v,\, u+w,\, v+w,\, u+v+w \}.
    \label{eq:seven-set}
\end{equation}
Any triple of linearly independent vectors from $\mathcal{S}$ generates the same subspace and represents the same term up to Clifford equivalence. Geometrically, the seven elements correspond to the points of the Fano plane, and the 28 valid triples to its triangles (Fig.~\ref{fig:flip-graph}a).

For a fixed vector $z$, let $\mathcal{T}_z = \{ q : z \in \mathcal{S}_q \}$ be the set of all terms whose seven-set contains $z$. The $\mathrm{GL}(3,2)$ symmetry allows choosing a representative triple $(u_q, v_q, z)$ for each such term. The combined contribution to $f_3$ then factorizes as $(z \cdot x)$ times a quadratic form:
\begin{equation}
    \sum_{q \in \mathcal{T}_z} (u_q \cdot x)(v_q \cdot x)
    = \sum_{i < j} B_{ij}\, x_i x_j,
\end{equation}
where equality is understood up to linear terms and $B$ is the alternating matrix
\begin{equation}
    B_{ij} = \sum_{q \in \mathcal{T}_z}
    (u_{qi}\, v_{qj} + u_{qj}\, v_{qi}).
    \label{eq:alternating-matrix}
\end{equation}

Any alternating matrix admits a canonical block-diagonal form via congruence $B \mapsto S\T B S$, computable by symplectic Gaussian elimination (SGE) in $O(n^3)$ time~\cite{witt_1937}. SGE performs simultaneous row and column additions (adding row $j$ to row $i$ together with column $j$ to column $i$), preserving skew-symmetry while reducing to canonical form. The result consists of $r$ blocks {\footnotesize $\bigl(\begin{smallmatrix} 0 & 1 \\ 1 & 0 \end{smallmatrix}\bigr)$} followed by zeros; each block corresponds to one $CCZ$ gate, so the $|\mathcal{T}_z|$ original terms reduce to $r$ terms. A worked example appears in the Supplementary Information.

The reduction is applied iteratively, selecting at each step a vector $z$ appearing in at least two seven-sets. In greedy mode, we choose the $z$ maximizing $|\mathcal{T}_z| - r$. In beam search mode, we maintain a pool of the top $k$ decompositions, expand each by all improving reductions, and retain the best $k$ results. All beam search experiments use width $k=2^{10}$. The search terminates when no choice of $z$ reduces the rank.

\subsection{Flip graph search} \label{sec:methods:d}

The flip graph at rank $r$ has vertices corresponding to CP decompositions of the target cubic polynomial with $r$ terms, connected by three local transformations (Fig.~\ref{fig:flip-graph}b). A \emph{flip} rewires two terms sharing a common factor $z$: given $(u_1, v_1, z)$ and $(u_2, v_2, z)$, it produces $(u_1 + u_2, v_1, z)$ and $(u_2, v_1 + v_2, z)$. Both configurations contribute the same cubic polynomial, but the new decomposition may expose shared factors that were previously unavailable, enabling further reductions. A \emph{reduction} applies SGE to all terms sharing a common factor, decreasing the rank. A \emph{plus} transition~\cite{kauers_2023} applies to two terms with disjoint seven-sets, replacing $u_1 v_1 w_1 + u_2 v_2 w_2$ by $(u_1 + u_2) v_1 w_1 + u_2 (v_1 + v_2) w_2 + u_2 v_1 (w_1 + w_2)$, temporarily increasing the rank to escape local plateaus.

The search proceeds via random walks, following~\cite{kauers_2023,moosbauer_2025,khoruzhii_2025}. We maintain a pool of $S = 10^3$ decompositions at the current rank~$r$. Each iteration selects a random decomposition and performs a random walk: at each step, a random flip is sampled from available candidates. If $P = 5 \times 10^4$ consecutive steps pass without a reduction, a plus transition is attempted. The walk terminates after at most $L = 10^6$ steps or upon reaching a lower rank. Successful walks contribute to the next-level pool at rank $r-1$; once this pool accumulates $S$ decompositions, the search advances and repeats.

Reductions are detected at two granularities. During flips, if two terms have seven-sets intersecting in 3 or 7 elements, the rank decreases by 1 or 2 respectively; such reductions are applied immediately. Full SGE scans over all shared factors are more expensive and performed every $R = 10^4$ steps.

Maintaining a diverse pool is essential: different decompositions at the same rank may occupy different regions of the flip graph, and reductions can lead to isolated components with no further descent paths (Fig.~\ref{fig:flip-graph}c). The plus transition bridges such components by temporarily increasing rank before finding alternative descent paths (see Supplementary Information). All experiments use the parameters above. Larger pools and longer walks may reach lower ranks at increased computational cost. Our implementation achieves approximately $10^6$ flip steps per second per thread.

\subsection{Bilinear circuits} \label{sec:methods:e}

A quantum circuit implementing a bilinear map performs
\begin{equation}
    \ket{a, b, c} \mapsto \ket{a, b, c \oplus g(a, b)},
\end{equation}
where $g_k(a,b)=\sum_{ij}G_{ijk}\,a_i b_j$. Each monomial $a_i b_j$ is accumulated into $c_k$ by one Toffoli gate. After expressing each Toffoli as Hadamard-conjugated $CCZ$, adjacent Hadamards cancel pairwise, leaving $H$ gates only at the circuit boundaries (Fig.~\ref{fig:gf-mult}). The remaining non-Clifford part implements $f_3 = \sum_{ijk} G_{ijk}\, a_i b_j c_k$, so minimizing the Toffoli count in this bilinear representation equals minimizing the CP rank of~$G$.

A CP term $(U_q,V_q,W_q)$ implements $c\leftarrow c+W_q(U_q\cdot a)(V_q\cdot b)$. This is a parity-controlled Toffoli (PCTOF) gate~\cite{khattar_2025}, so a decomposition with $r$ terms gives $r$ PCTOF gates. Each is realizable by one Toffoli conjugated by CNOT circuits within the three registers.

Unlike the cubic phase polynomial setting, the tensor $G$ has dimensions $(n_a, n_b, n_c)$ and the $\mathrm{GL}(3,2)$ symmetry does not apply. This is the standard bilinear setting of the original flip graph framework~\cite{kauers_2023}: flips operate on pairs sharing a common factor along any of the three axes, plus transitions introduce a third term from two terms with no shared factors, and reductions occur when terms with a shared factor have linearly dependent vectors along another axis. We use the same FGS parameters as in the cubic setting.

For finite field multiplication over $\mathrm{GF}(2^p)$, elements are polynomials $a(x) = \sum_{k=0}^{p-1} a_k x^k$ with binary coefficients, and multiplication is performed modulo an irreducible polynomial $h(x)$ of degree~$p$. The circuit decomposes into two stages. Polynomial multiplication computes $a(x)\,b(x)$ via the convolution $(ab)_\ell = \sum_{i+j=\ell} a_i b_j$, defining the convolution tensor $G^{\mathrm{conv}} \in \F_2^{p \times p \times (2p-1)}$ with $G^{\mathrm{conv}}_{ij\ell} = 1$ iff $i + j = \ell$. Modular reduction $ab \bmod h$ is a linear map $R\colon \F_2^{2p-1} \to \F_2^{p}$ requiring no additional multiplications. Their composition defines the field tensor $G^{\mathrm{GF}} \in \F_2^{p \times p \times p}$:
\begin{equation}
    G^{\mathrm{GF}}_{ijk} = \sum_{\ell} G^{\mathrm{conv}}_{ij\ell}\, R_{k\ell}.
\end{equation}

Any rank-$r$ decomposition of $G^{\mathrm{conv}}$ induces a rank-$r$ decomposition of $G^{\mathrm{GF}}$ by projecting $W'_{kq} = R_{kl}\, W_{lq}$. A decomposition of $G^{\mathrm{GF}}$ embeds into a phase polynomial circuit via $(i, j, k) \mapsto (i,\, p+j,\, 2p+k)$, mapping the $p \times p \times p$ tensor to a $3p \times 3p \times 3p$ tensor with factors supported on disjoint index blocks corresponding to the $a$, $b$, $c$ registers. We optimize both tensors and report the best result in Table~\ref{tab:gf-mult}.

Converting $Z$, $CZ$, and $CCZ$ gates back to $X$, CNOT, and Toffoli gates by Hadamard conjugation leaves no $H$ gates in the bilinear circuit. The Hadamards are absorbed into the individual gates using $H^2=I$ and
\begin{equation}
    H_a H_b\,\mathrm{CNOT}_{ab}\,H_a H_b = \mathrm{CNOT}_{ba}.
\end{equation}
The phase representation thus serves as an intermediate step between the original $\{X,\mathrm{CNOT},\mathrm{Toffoli}\}$ circuit and an optimized circuit over the same gate set.

\begin{table}
    \caption{\justifying
    \textbf{Gate counts for GF$(2^p)$ multiplication circuits.}
    The circuits implement $\ket{a,b,c}\mapsto\ket{a,b,c+ab}$ over the finite field $\mathrm{GF}(2^p)$, with arbitrary accumulator input $c$.
    The Polynomial column gives the irreducible polynomial in $t$ used to represent each field.
    The tensor decompositions are based on the constructions of~\cite{cenk_2010}. Cubic-phase routing (CPR) constructs ancilla-free $\{\mathrm{CNOT},\mathrm{Toffoli}\}$ circuits on $3p$ qubits.
    }
    \label{tab:gf-mult-cnot-short}
    \centering
    \begin{tabular}{l@{\hspace{1.5em}}l@{\hspace{1.5em}}r@{\hspace{1.5em}}r}
        \toprule
        Circuit & Polynomial & Toffoli & CNOT \\
        \midrule
        GF$(2^2)$ Mult     & $t^2+t+1$                       &   3 &      6 \\
        GF$(2^3)$ Mult     & $t^3+t+1$                       &   6 &     16 \\
        GF$(2^4)$ Mult     & $t^4+t+1$                       &   9 &     26 \\
        GF$(2^5)$ Mult     & $t^5+t^2+1$                     &  13 &     42 \\
        GF$(2^6)$ Mult     & $t^6+t+1$                       &  15 &     70 \\
        GF$(2^7)$ Mult     & $t^7+t+1$                       &  22 &     98 \\
        GF$(2^8)$ Mult     & $t^8+t^4+t^3+t+1$               &  24 &    137 \\
        GF$(2^9)$ Mult     & $t^9+t+1$                       &  30 &    185 \\
        GF$(2^{10})$ Mult  & $t^{10}+t^3+1$                  &  33 &    260 \\
        GF$(2^{12})$ Mult  & $t^{12}+t^3+1$                  &  42 &    308 \\
        GF$(2^{16})$ Mult  & $t^{16}+t^5+t^3+t+1$            &  60 &    690 \\
        GF$(2^{32})$ Mult  & $t^{32}+t^7+t^3+t^2+1$          & 135 &  3\,078 \\
        GF$(2^{64})$ Mult  & $t^{64}+t^4+t^3+t+1$            & 300 & 15\,358 \\
        GF$(2^{128})$ Mult & $t^{128}+t^7+t^2+t+1$           & 672 & 68\,571 \\
        \bottomrule
    \end{tabular}
\end{table}

\subsection{Cubic-phase routing} \label{sec:methods:f}

After Toffoli-count minimization, a CP decomposition fixes the cubic phase $(-1)^{f_3(x)}$ but leaves freedom in its CNOT realization. Cubic-phase routing (CPR) searches for a CNOT route that realizes these terms with $CCZ$ gates. The resulting lower-degree phase is realized by QPR. We count each $CZ$ gate as one CNOT, treat single-qubit Clifford gates as free, and assume all-to-all connectivity.

Let $b_q\in\F_2^n$ denote the parity carried by qubit~$q$, and let $b_{qj}\in\mathrm{GL}(n,2)$ be the matrix with rows $b_q$. Initially, $b_{qj} = \delta_{qj}$. A CNOT from qubit~$q'$ to qubit~$q$ updates $b_q\leftarrow b_q+b_{q'}$. We call a parity resident when it occurs as a row of~$b_{qj}$. To load a parity $u=\sum_q\alpha_qb_q$, any row~$q$ with $\alpha_q=1$ can serve as the target. Adding the remaining selected rows into~$q$ loads $u$ using $\operatorname{nnz}(\alpha)-1$ CNOT gates~\cite{amy_2019a}. Each search step prepares the three parities of one unfinished term and applies its $CCZ$ gate; resident parities are reused without loading.

Each cubic term is defined up to the $\mathrm{GL}(3,2)$ symmetry: any independent triple from its seven-set has the same cubic part and differs only by a lower-degree Clifford phase. For each unfinished term, CPR chooses the triple with minimum total coordinate weight in the current basis among the 28 representatives of its seven-set.

From each state, we consider every unfinished term. Its three factors are loaded in increasing order of their coordinate weights at the start of the step. For each load, we consider every compatible target except rows already holding another factor of the selected triple. After each parity, a local beam retains up to four partial loading sequences with the lowest CNOT counts. The resulting children each complete the selected term, leaving the loaded parities available to later terms.

A beam state records the current basis, the completed terms, and the routed CNOT count. All states at a given search depth have completed the same number of terms. Across all candidate terms and their loading sequences, we retain up to $k$ states ranked by
\begin{equation}
    \text{routed CNOT count}+|\{q:b_q\ne e_q\}|,
\end{equation}
where $e_q$ is the $q$th coordinate vector. The number of nonidentity rows $|\{q:b_q\ne e_q\}|$ is a lower bound on the CNOT cost of restoring the basis, since each such row must be targeted at least once. For the same basis and completed-term set, we retain only the route with the lowest CNOT count.

After $\sub{r}{cp}$ steps, every retained route has applied all terms. We select the route minimizing the routed CNOT count plus $\mathrm{PMH}(b_{qj})$, where $\mathrm{PMH}(b_{qj})$ is the CNOT count of a Patel--Markov--Hayes realization of the current basis~\cite{patel_2008}. Its basis is restored by greedy-PMH. At each greedy step, we inspect all $n(n-1)$ possible CNOTs and select the one minimizing the PMH completion count, provided that the new count plus one is smaller than the current count. When no improvement remains, PMH completes the residual map.

For bilinear circuits, the three factors occupy disjoint registers, so their bases are routed separately and the $\mathrm{GL}(3,2)$ symmetry is not used. The same term-by-term search prepares the factors register by register. The selected $\mathrm{GF}(2^p)$ decompositions in Table~\ref{tab:gf-mult-cnot-short} have $U=V$, allowing one input route to be mirrored on both registers. The disjoint supports induce no lower-degree phase, and Hadamard conjugation gives an ancilla-free $\{\mathrm{CNOT},\mathrm{Toffoli}\}$ accumulator circuit. The cubic and finite-field benchmarks use beam widths $k=32$ and $k=1024$, respectively.

\subsection{Quadratic-phase routing} \label{sec:methods:g}

CPR fixes a \{CNOT, $CCZ$\} route for the cubic phase, but a lower-degree Clifford phase $(-1)^{\ell(x)+Q(x)}$ may remain, where $\ell$ is linear and $Q$ is homogeneous quadratic. This phase can be part of the target or induced by the selected cubic representatives. Quadratic-phase routing (QPR) inserts \{$CZ$, $S$, $X$\} gates into the fixed route to realize~$Q$ with few additional two-qubit gates.

Some quadratic phases can be absorbed without additional two-qubit gates. Inserting $S$ gates around a CNOT gives~\cite{maslov_2023}
\begin{equation}
    S_a S_b^\dagger\,\mathrm{CNOT}_{a\to b}\,S_b\ket{a,b}=(-1)^{ab}\ket{a,a+b},
\end{equation}
while inserting an $X$-gate pair around a $CCZ$ gives
\begin{equation}
    X_a CCZ_{abc}X_a\ket{a,b,c}=(-1)^{abc+bc}\ket{a,b,c}.
\end{equation}
The analogous insertions on the other two qubits add $ac$ and $ab$. The homogeneous quadratic forms $f_i$ generated by the available $S$ and $X$ insertions span a space~$F$. It is therefore sufficient to realize~$Q$ mod~$F$. The remaining element of~$F$ is implemented by the corresponding one-qubit gate insertions.

A $CZ$ gate between qubits carrying parities $u$ and $v$ realizes the homogeneous quadratic part of~$uv$. After basis restoration, all coordinate forms $x_i x_j$ with $i<j$ are available and already give a complete dictionary. The full dictionary additionally uses the parities occurring along the route. After a CNOT from $c$ to $t$, the updated target carries $b_t\leftarrow b_t+b_c$; all newly products therefore have the form $b_tb_v$ with $v\ne t$. Products of Boolean linear forms can contain diagonal terms $x^2=x$; these are added to~$\ell$.

Let $d_j$ denote the additional path forms. We represent a realization as
\begin{equation}
    Q=R+\sum_i s_i f_i+\sum_j t_j d_j,
\end{equation}
where $s$ and $t$ are binary vectors selecting free and path insertions, and $R$ is the residual quadratic polynomial in the input coordinates. Each monomial of~$R$ requires one $CZ$ after basis restoration, so the cost is $\operatorname{nnz}(t)+\operatorname{nnz}(R)$. In the quotient by~$F$, minimizing this cost is minimum-weight syndrome decoding for a binary linear code, which is NP-hard in general~\cite{berlekamp_1978}.

Gaussian elimination on the free forms gives a representative of~$Q$ modulo~$F$, together with the corresponding free insertion coefficients. We start with no path insertions and the cheaper of this representation and the direct coordinate realization $R=Q$, $s=0$. If $Q\in F$, this already gives zero cost.

We then use local descent in the spirit of gradient-like decoding~\cite{ashikhmin_1998}. Each pass scans the original free forms, followed by the path forms. A free move toggles $s_i$ and updates $R\leftarrow R+f_i$; a path move toggles $t_j$ and updates $R\leftarrow R+d_j$. We accept a move only if it strictly decreases the cost and repeat the passes until no improvement remains. The result is a local minimum with respect to these moves and costs at most $\operatorname{nnz}(Q)$. The selected coefficients directly determine the $S$, $X$, and path $CZ$ insertions, while~$R$ gives the terminal $CZ$ gates. The accumulated linear phase is realized by $Z$ gates.

\begin{table}
    \caption{\justifying
    \textbf{Gate counts for 20-qubit XOR AES S-box circuits.}
    All circuits implement $\ket{x,y,0^4}\mapsto\ket{x,y+S(x),0^4}$, where $S$ is the Advanced Encryption Standard (AES) S-box. The registers $x,y$ contain eight qubits each; the four clean workspace qubits are restored to zero.
    $^*$Block-structured circuit shown in Fig.~\ref{fig:aes-sbox-building-blocks}d.
    $^\dag$Manually optimized for CNOT count.
    $^\ddag$Manually optimized for Toffoli depth.
    }
    \label{tab:aes-sbox-cnot}
    \centering
    \begin{tabular}{lr@{\hspace{1.5em}}r@{\hspace{1.5em}}r}
        \toprule
        Construction & Toffoli & CNOT & Toffoli depth \\
        \midrule
        Li et al.~\cite{li_2023} & 54 & 238 & 42 \\
        Chen et al.~\cite{chen_2025b} & 52 & 226 & 32 \\
        \midrule
        This work$^*$ & 52 & 186 & 44 \\
        This work$^\dag$ & 52 & 147 & 47 \\
        This work$^\ddag$ & 52 & 184 & 24 \\
        \bottomrule
    \end{tabular}
\end{table}

\subsection{Implementation details} \label{sec:methods:h}

The benchmark collection was used in earlier circuit-optimization studies~\cite{amy_2014,heyfron_2018} and is distributed in the Feynman repository~\cite{amy_2019b}. For the cubic benchmark circuits in Table~\ref{tab:a-bench} that were optimized without splitting in~\cite{ruiz_2025}, we used their tensors directly, retaining the circuit preparation and Hadamard gadgetization of that work. We also use their tensors for all 35 circuits in the extended benchmark suite. We use the input circuits and Clifford\,+\,Phase\,+\,Clifford compilation procedure from~\cite{vandaele_2025a} to obtain a second set of tensors for Table~\ref{tab:a-bench}, including the remaining seven cubic circuits and QFT$_4$. The prior counts in Table~\ref{tab:a-bench} are taken from~\cite{ruiz_2025,vandaele_2025a} and include results obtained earlier in~\cite{heyfron_2018,beaudrap_2020,beaudrap_2020a,vandaele_2025}.

This compilation applies FastTMerge~\cite{vandaele_2024a}, followed by InternalHOpt~\cite{vandaele_2024}, before gadgetizing the remaining internal Hadamard gates. The reported $h$ counts refer to internal Hadamard gates after the corresponding circuit preprocessing, not in the original circuit. The $h$ and Size columns refer to the tensor representation used for the reported Toffoli counts. 

For Toffoli minimization, the two compilation approaches give the same best counts found on all circuits where both apply. For $T$-count minimization, the results can differ; we report the best across both compilations.

For the cubic and finite-field benchmarks, CNOT optimization starts from one fixed decomposition per circuit. Its factors are expressed in the original phase-block coordinates, incorporating any basis change made during BCO. The finite-field circuits in Table~\ref{tab:gf-mult-cnot-short} use decompositions based on~\cite{cenk_2010}, rather than those obtained by the automatic search in Table~\ref{tab:gf-mult}. The linear blocks in Fig.~\ref{fig:aes-sbox-building-blocks}d were constructed by beam search over layers of CNOTs and SWAPs, with SWAPs treated as free wire relabellings.

Computations were performed on a Intel Xeon Gold 6342 processor (24 cores, 48 threads) with a 24-hour time limit per search run.

\subsection{Code availability}

The source code for this study is available at 
\href{https://github.com/ZIB-IOL/polytof}{\texttt{github.com/ZIB-IOL/polytof}}. 
The C++ implementation is self-contained with no external dependencies 
and includes verification scripts and runnable examples to reproduce 
all reported results.

\section*{Data Availability}

All circuits, decompositions, and datasets generated during this study 
are available in the public repository 
\href{https://github.com/ZIB-IOL/polytof}{\texttt{github.com/ZIB-IOL/polytof}}.

\section*{Acknowledgments}

The authors acknowledge the support of the DFG Cluster of Excellence MATH+ (EXC-2046/2, project id 390685689) funded by the Deutsche Forschungsgemeinschaft (DFG), and the National High-Performance Computing (NHR) network.

\section*{Funding}

This research received institutional support from the Zuse Institute Berlin (ZIB). Beyond this institutional support, the research and preparation of the manuscript were carried out by the authors.

\section*{Author Contributions}

K.K. developed the methods, implemented the software, and performed the computations. P.G. and S.P. supervised the research. All authors provided critical feedback, discussed the results, and revised the manuscript.

\section*{Competing Interests}

The authors declare no competing financial or non-financial interests.

\bibliographystyle{naturemag}
\hypersetup{hidelinks}
\bibliography{references}

\end{document}